\documentclass[pdflatex,sn-mathphys]{sn-jnl}

\usepackage{graphicx}
\usepackage{amsmath}
\usepackage{bm}
\usepackage{amsfonts}
\usepackage{amssymb}
\usepackage{hyperref}
\usepackage{multirow}
\usepackage{subfigure}

\usepackage[normalem]{ulem}

\def\br{\bm{\rho}}
\def\de{\mathrm{d}}

\begin{document}

\title[Periodic patterns for resolution limit characterization of CPI]{Periodic patterns for resolution limit characterization of correlation plenoptic imaging}


\author[1,2]{\fnm{Francesco} \sur{Scattarella}}

\author[1,2]{\fnm{Gianlorenzo} \sur{Massaro}}

\author[3]{\fnm{Bohumil} \sur{Stoklasa}}

\author*[1,2]{\fnm{Milena} \sur{D'Angelo}}\email{milena.dangelo@uniba.it}

\author[1,2]{\fnm{Francesco V.} \sur{Pepe}}

\affil[1]{\orgdiv{Dipartimento Interateneo di Fisica}, \orgname{Universit\`a degli Studi di Bari}, \orgaddress{\postcode{I-70126}, \city{Bari}, \country{Italy}}}

\affil[2]{\orgname{Istituto Nazionale di Fisica Nucleare}, \orgdiv{Sezione di Bari}, \orgaddress{\city{Bari}, \postcode{I-70125}, \country{Italy}}}

\affil[3]{\orgdiv{Department of Optics}, \orgname{Palack\'y University}, \orgaddress{\postcode{77146}, \city{Olomouc}, \country{Czech Republic}}}

\abstract{The measurement of the spatio-temporal correlations of light provides an interesting tool to overcome the traditional limitations of standard imaging, such as the strong trade-off between spatial resolution and depth of field. In particular, using correlation plenoptic imaging, one can detect both the spatial distribution and the direction of light in a scene, pushing both resolution and depth of field to the fundamental limit imposed by wave-optics. This allows one to perform refocusing of different axial planes and three-dimensional reconstruction without any spatial scanning. In the present work, we investigate the resolution limit in a particular correlation plenoptic imaging scheme, by considering periodic test patterns, which provide, through analytical results, a deeper insight in the resolution properties of this second-order imaging technique, also in comparison with standard imaging.}


\maketitle

\section{Introduction}\label{sec:introduction}

Plenoptic imaging (PI) refers to a class of devices and techniques that can detect the \textit{light field}, \textit{i.e.} the combined information on the spatial distribution and propagation direction of light, in a single exposure of a scene of interest \cite{adelson}. PI is being used in a variety of applications, including microscopy \cite{microscopy1,microscopy2,microscopy3,microscopy4}, stereoscopy \cite{adelson,muenzel,levoy}, wavefront sensing \cite{thesis_wu,eye,atmosphere1,atmosphere2}, particle image velocimetry \cite{piv}, particle tracking and sizing \cite{tracking}, and photography, where it is used to add refocusing capabilities to digital cameras \cite{ng}. Moreover, 3D neuronal activity functional imaging \cite{microscopy4}, surgery \cite{surgery}, endoscopy \cite{endoscopy}, and flow visualization \cite{piv2} are examples of cutting-edge applications. A state-of-the-art plenoptic device offers the possibility to perform, with a single acquisition and after numerical data treatment, various applications such as changing viewpoint, refocusing at different depths and, consequently, 3D reconstruction of the scene \cite{3dimaging}. This is performed thanks to three optical elements, which are the basis of a plenoptic camera: main lens, an array of micro-lenses and a detector. Nevertheless, the lateral resolution of focused images obtained with such a device is essentially limited by the micro-lens size, making the diffraction-limited resolution set by the numerical aperture of the main lens unreachable. 

A way to overcome this practical limitation is offered by a new technique, named correlation plenoptic imaging (CPI), capable of performing plenoptic imaging without spatial resolution loss \cite{cpi_prl,cpi_exp} and reaching the diffraction limit. This is done by measuring second-order spatio-temporal correlations of light: thanks to the correlated nature of light beams \cite{cpi_prl,cpi_qmqm,cpi_technologies,cpi_jopt}, such a measurement encodes information not only on the spatial distribution of light in a given plane in the scene (as in ghost imaging implementations \cite{PittShi,gatti,laserphys,valencia,scarcelliPRL}), but also on the direction of light. This method mitigates significantly the spatial vs. directional resolution trade-off, paving the way toward the development of novel quantum plenoptic cameras, which will enable one to perform the same tasks of standard plenoptic systems, such as refocusing and scanning-free 3D imaging, along with a relevant performance increase in terms of resolution (which can be diffraction limited), depth of field (DOF), and noise \cite{qu3d}.

CPI involves coherent properties of light, as demonstrated in \cite{cpi_prl}. Therefore, an unambiguous definition of resolution limit for all CPI protocols is not trivial as for incoherent imaging: the standard criteria, based on the properties of a point-spread function, become meaningless in CPI. This also lead to the ambiguity about the type of sample that should be used to accurately evaluate the CPI resolution limit.
In a previous work, we have analyzed the resolution limit of CPI using the paradigmatic case of an object with Gaussian profile object \cite{resol_limit}, and worked in what we consider as the most flexible CPI scheme: correlation plenoptic imaging \textit{between arbitrary planes} (CPI-AP) \cite{cpiap}. Unlike other CPI protocols, this scheme only requires a single focusing element (see, e.g., its experimental implementation in Ref. \cite{CPIAPexp}). 
In the present paper, we extend the results obtained in Ref. \cite{resol_limit} by considering an object with a periodic profile, which is a conventional object used for characterizing the resolution of traditional optical systems, for instance, in terms of the modulation trasnfer function (MTF) \cite{howland1983new}; a periodic object is indeed useful ahead of the decomposition of any object into space harmonics. We achieve analytical results that offer the possibility to directly compare CPI with standard imaging based on direct intensity measurement, both qualitatively and quantitatively.

\section{Methods}\label{sec:methods}

In CPI, light from a chaotic source is split in two beams that propagate along two distinct optical paths $a$ and $b$. Correlation between intensities measured at the transverse coordinates $\br_{a}=(x_a,y_a)$ and $\br_b=(x_b,y_b)$, respectively placed on detectors $D_a$ and $D_b$ at the end of each path, encodes more information than the average intensities. Specifically, the relevant imaging properties are encoded in the correlation between intensity fluctuations
\begin{equation}\label{Gammagen}
\Gamma(\br_a,\br_b) = \langle \Delta I_a(\br_a) \Delta I_b(\br_b) \rangle = \langle I_a(\br_a) I_b(\br_b) \rangle - \langle I_a(\br_a) \rangle \langle I_b(\br_b) \rangle .
\end{equation}

\begin{figure}
    \centering
    \includegraphics[width=0.8\textwidth]{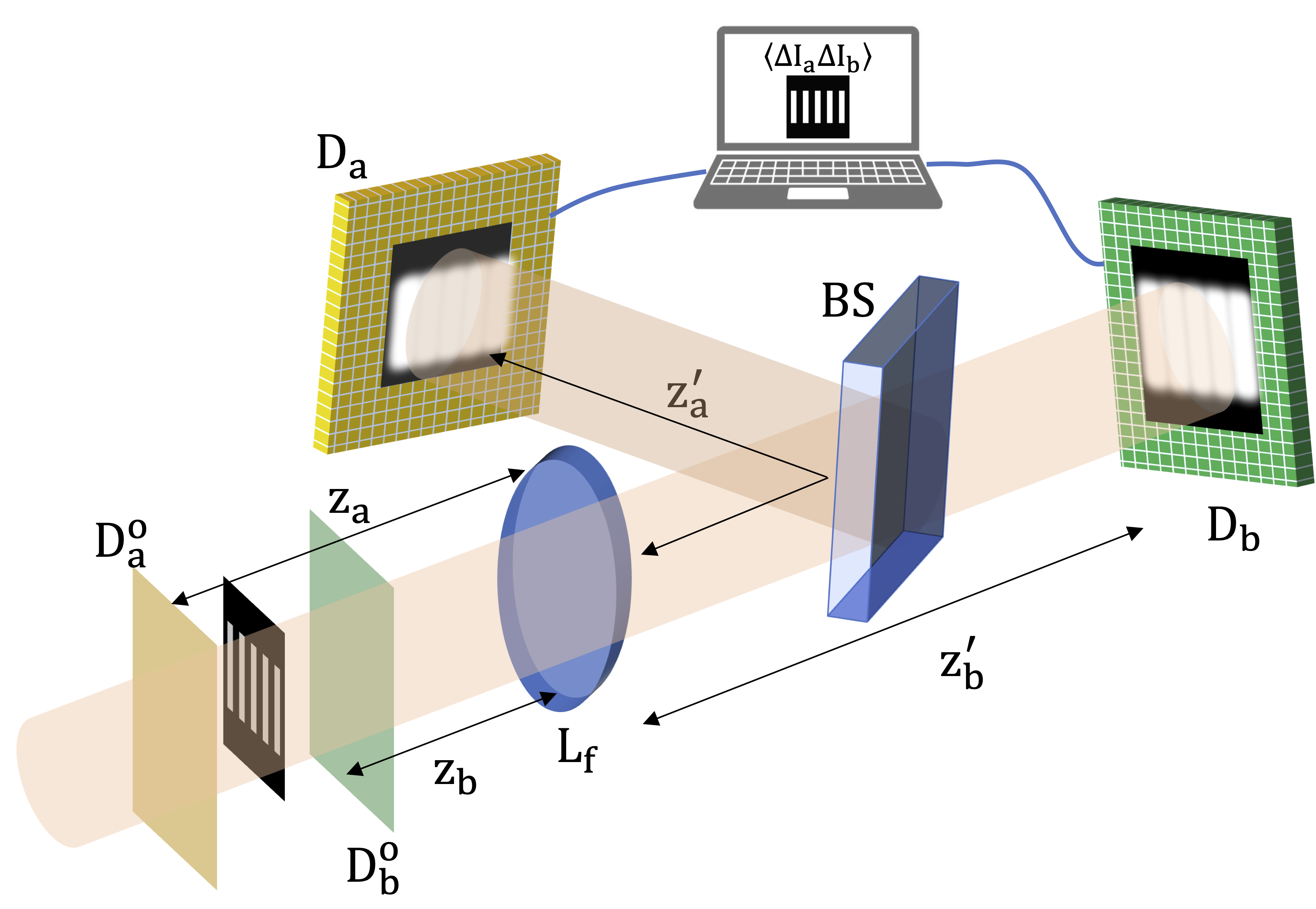}
    \caption{Optical scheme of correlation plenoptic imaging between arbitrary planes (CPI-AP) \cite{cpiap}. The object is assumed to be a chaotic light emitter. The lens $L_f$ focuses the images of the two planes $D^o_a$ and $D^o_b$ on the two spatially-resolving sensors $D_a$ and $D_b$, respectively. Information on the distribution and direction of light from the object is collected by correlating the intensity fluctuations retrieved by each pair of pixels, one on each detector. }
    \label{fig:setup}
\end{figure}

A typical CPI-AP experimental scheme is shown in Figure~\ref{fig:setup} \cite{cpiap}. Light comes from an object, which emits chaotic light, propagates toward the lens $L_{f}$ with focal length $f$, and then encounters a beam splitter (BS). The latter generates two copies of the input beam, which are eventually detected by the two sensors $D_{a}$ and $D_{b}$, both spatially resolving. The detectors are placed in such a way to collect the focused images of two planes in proximity of the object, called $D^{o}_{a}$ and $D^{o}_{a}$.
As demonstrated in \cite{cpiap}, plenoptic information can be retrieved by analyzing the spatio-temporal correlations between the fluctuations in the intensity acquired by the two sensors.
For evaluating $\Gamma(\br_a,\br_b)$ in the discussed CPI-AP setup, we shall assume that the object is positioned at a distance $z$ from the lens $L_{f}$. Light emitted by this object is characterized by the intensity profile $A(\br_{o})=A(x_o,y_o)$. We further assume that transverse coherence can be safely neglected, and that emission is quasi-monochromatic around the central wavelength $\lambda$ (corresponding to the wavenumber $k=2\pi/\lambda$). Neglecting irrelevant factors (independent of $\br_a$ and $\br_b$), the resulting correlation function reads

\begin{equation}\label{GammaAP}
    \Gamma(\br_a,\br_b) = \left\lvert \int \de^2\br_o A(\br_o) g_a^*(\br_a,\br_o) g_b(\br_b,\br_o) \right\rvert^2 ,
\end{equation}

where
\begin{equation}\label{propagator}
 g_j(\br_j,\br_o)= \int \de^2\br_{\ell} P(\br_{\ell}) \, \text{exp}\left\{-ik\left[\left(\frac{1}{z}-\frac{1}{z_{j}}\right)\frac{\br^{2}_{\ell}}{2}-\left(\frac{\br_{o}}{z}-\frac{\br_{j}}{M_{j}z_{j}}\right)\cdot \br_{\ell}\right]\right\},
\end{equation}
with $j = a,b$, are the paraxial optical transfer functions \cite{goodman} along the two paths, $P(\br_{l})$ is the pupil function of the lens $L_f$ and $M_{j}$ are the magnifications of the object planes $D^{o}_{j}$ on detectors $D_{j}$.

The plenoptic properties of $\Gamma(\br_{a},\br_{b})$ from Eq.~\eqref{propagator} can be fully understood by considering the dominant contribution to the integrals in the limit $k\rightarrow\infty$ of geometrical optics. In general, $\Gamma(\br_{a},\br_{b})$ encodes the image of the squared intensity profile $A^2(\br_0)$ of an object, generally depending on both detector coordinates $\br_a$ and $\br_b$, unless $z=z_a$ or $z=z_b$ (i.e., the object is focused on $D_a$ or $D_b$, respectively). In the other cases,
proper linear combinations of the detector coordinates can be defined \cite{refocus}
\begin{equation}\label{rho_rs}
\br_r= \frac{1}{z_b-z_a} \left(\frac{z-z_a}{M_b}\br_b - \frac{z-z_b}{M_a}\br_a \right),\quad
\br_s= \frac{1}{z_b-z_a} \left(\frac{z_b}{M_a}\br_a - \frac{z_a}{M_b}\br_b \right),
\end{equation}
in such a way that the ``refocused'' correlation function

\begin{align}\label{Gammarefo}
    \Gamma_{\mathrm{ref}}(\br_r,\br_s) & :=\Gamma\left[ M_a \left( \frac{z_a}{z} \br_r + \left( 1 - \frac{z_a}{z}\right)\br_s \right), M_b \left( \frac{z_b}{z} \br_r + \left( 1 - \frac{z_b}{z}\right)\br_s \right) \right] \nonumber \\ & \sim A^2\left(\br_r\right)\left\lvert P\left(\br_s\right)\right\rvert^4 ,
\end{align}

encodes the product of a squared image of the object, depending only on $\br_r$, and a squared image of the pupil intensity transmission profile, depending only on $\br_s$. The function $\Gamma_\mathrm{ref}$ can now be integrated over $\br_s$ variable, to obtain the \textit{refocused image}
\begin{equation}\label{refocus}
\Sigma_{\mathrm{ref}}(\br_r) = \int \de^2\br_s \Gamma_\mathrm{ref}\left(\br_r, \br_s\right) \sim A^2(\br_r).
\end{equation}
Notice that, the geometrical optics approximation tends to become exact in the limit $k\to\infty$ (namely, in the case of wavelength much smaller than the object details), and the refocused image tends to coincide with the squared object intensity profile. 

Due to the structure of the correlation function of Eq.~\eqref{GammaAP}, the refocused image of Eq.~\eqref{refocus} takes the form
\begin{equation}
    \Sigma_{\mathrm{ref}}(\br_r) = \int \de^2\br_o A(\br_o) \int \de^2\br'_o A(\br'_o) \Phi(\br_r;\br_o,\br'_o)
\end{equation}
namely, it is a double integral on the object coordinates, with $\Phi$ a function involving the optical propagators. Such a feature prevents us to define a proper point-spread function, as one naturally does in the case of the direct intensity image. Actually, as one can observe from the above equation, the object itself appears in the second integral, which determines the image spread: such a general feature will be evidently encountered even in the special case that we consider in this work. The definition of the CPI resolution limit thus requires the choice of a specific class of test objects with finite detail size.

As a testbed, we consider a class of objects whose intensity profile is characterized by a periodic pattern in the $x$ direction
\begin{equation}\label{apert}
A(x_o,y_o) = A_0 \cos^2\left({\frac{\pi}{d} x_o}\right) 
\end{equation}
of spatial frequency $\pi/d$, with $d$ the peak-to-peak distance. For example, periodic gratings can be considered objects of this type. This  allows to perform an analytical determination of the refocusing function and a direct comparison of results obtained in different cases. The lens aperture will be modelled by the Gaussian pupil function
\begin{equation}\label{lensaperture}
    P(\br_{\ell})= P_0 \exp\left( - \frac{\br_{\ell}^2}{2 \sigma^2}\right) = P_0 \exp\left( - \frac{x_{\ell}^2}{2 \sigma^2}\right) \exp\left( - \frac{y_{\ell}^2}{2 \sigma^2}\right) ,
\end{equation}
which has the advantage of being factorized in the two coordinates, providing at the same time a qualitatively convenient approximation to the actual lens pupil.

\section{Results}\label{sec:results}
Considering the periodic test objects \eqref{apert} and the assumption of Gaussian lens aperture \eqref{lensaperture}, we first compute the correlation function \eqref{GammaAP},
then, following the refocusing procedure defined by Eqs.~\eqref{rho_rs}--\eqref{refocus}, we obtain the periodic refocused image
\begin{equation} \label{sigmaref}
    \Sigma_{\mathrm{ref}}(x_r,y_r;d) = F(d) \left[ C_0(d)+C_1(d)\cos\left(\frac{2\pi}{d} x_r\right) + C_2(d) \cos\left(\frac{4\pi}{d} x_r\right) \right]
\end{equation}
where $F(d)$ is an irrelevant overall factor, and the coefficients $C_0(d)$, $C_1(d)$ and $C_2(d)$ are functions of the spatial periodicity of the object
\begin{align}
    C_0(d) = & \frac{1}{8} 
    \left\{ 2 + \exp\left[ - \frac{1}{2} \left( \frac{2 \pi z}{d k \sigma}\right)^2 \right] \right\} , \\
    C_1(d) = & \frac{1}{2} \exp\left[ - \frac{1}{8} \left( \frac{2 \pi z}{d k \sigma}\right)^2 \,\frac{3(\zeta_a^2+\zeta_b^2)-2\zeta_a\zeta_b}{(\zeta_a-\zeta_b)^2} \right] \cos \left( \frac{2\pi^2 z}{d^2 k} \frac{\zeta_a\zeta_b}{\zeta_a-\zeta_b} \right) \\ 
   C_2(d) = & \frac{1}{8} \exp\left[- \left( \frac{2 \pi z}{d k \sigma}\right)^2 \,\frac{\zeta_a^2 + \zeta_b^2}{(\zeta_a-\zeta_b)^2} \right] ,
\end{align}
where the dimensionless parameters
\begin{equation}
    \zeta_a = 1- \frac{z}{z_a}, \quad \zeta_b = 1- \frac{z}{z_b} , 
\end{equation}
are conveniently introduced.

\begin{figure}
\centering
\subfigure[]{\includegraphics[width=0.46\textwidth]{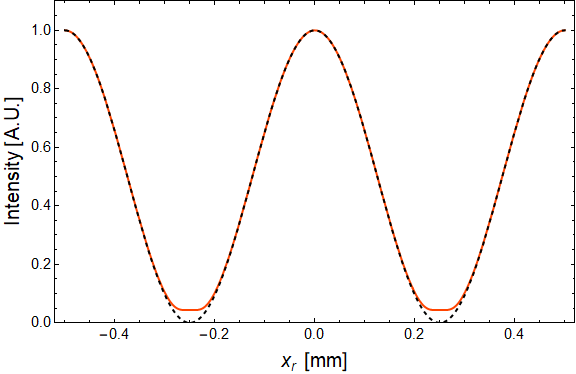}} \\
\subfigure[]{\includegraphics[width=0.46\textwidth]{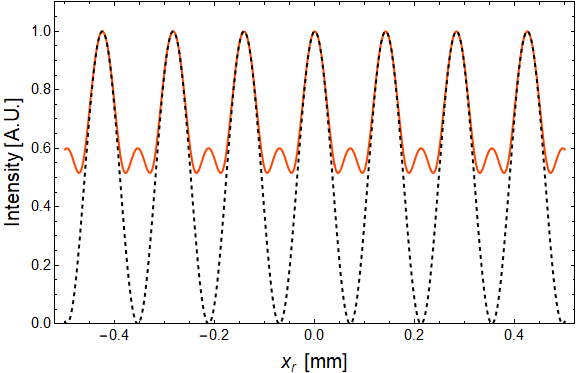}} \\
\subfigure[]{\includegraphics[width=0.46\textwidth]{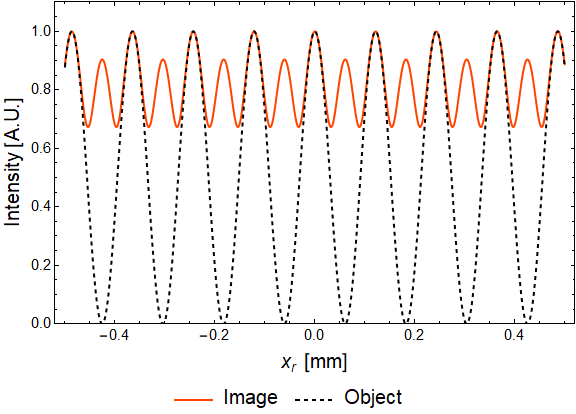}}
\caption{Comparison between periodic object profiles with different period $d$ and their square-rooted CPI-AP refocused images in three different cases: full (top panel, $d=0.50\,\mathrm{mm}$), intermediate (middle panel, $d=0.14\,\mathrm{mm}$), and low visibility (bottom panel, $d=0.12\,\mathrm{mm}$). Notice that, in panel (c), the refocused image shows secondary peaks due to the nonlinear structure of the correlation function; the class of resolution criteria defined in Eq.~\eqref{eq:resolution_condition} rules out the possibility for such an image to be considered ``resolved'' in any case. Numerical values of the parameters are reported in the text.\label{fig:object}}
\end{figure}

In the limit $k\to\infty$, Eq.~\eqref{sigmaref} tends to the exact expression for any values of object periodicity $d$, as expected from Eq.~\eqref{refocus}, providing a \textit{perfectly resolved image}:
\begin{equation} \label{sigmaref_exact}
    \Sigma_{\mathrm{ref}}(x_r,y_r;d) \sim \frac{3}{8}+\frac{1}{2}\cos\left(\frac{2\pi}{d} x_r\right)+\frac{1}{8}\cos\left(\frac{4\pi}{d} x_r\right) = \left[ \cos^2 \left( \frac{\pi}{d} x_r \right)\right]^2 .
\end{equation}
We can observe that the coefficients $C_{j}(d)$ are determined by factors such as the wavelength, the lens size $\sigma$, the distances $z_j$ of the two reference planes from the lens, and the object axial position $z$. It is worth noticing that in the finite-resolution condition these factors depend also on the object features (in this case, the periodicity $d$). In Figure \ref{fig:object}, we compare periodic object profiles with different $d$ and their refocused images. In all plots, numerical parameters are fixed to $\lambda=532\,\mathrm{nm}$, $z_a=293\,\mathrm{mm}$, $z_b=345\,\mathrm{mm}$, $\sigma=8.2\,\mathrm{mm}$, and the object axial position $z=z_m=(z_a+z_b)/2$. Notice that the width of the Gaussian pupil function corresponds to an effective lens diameter close to $30\,\mathrm{mm}$ \cite{cpi_exp}; hence, the numerical aperture of the lens, seen from the mid-plane $z_m=(z_a+z_b)/2$ is around $1/20$.
The top plot shows a refocused image (orange dashed line) almost perfectly overlapping with the object (black line), while the the middle and bottom plots represent cases of reduced visibility. In particular, the bottom plot shows a second-order interference effect that hinders the correspondence between the input pattern and its image.

\begin{figure}
\centering
(a)\includegraphics[width=0.45\textwidth]{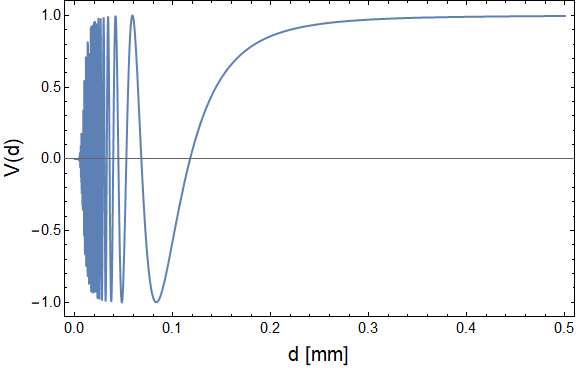}
(b)\includegraphics[width=0.45\textwidth]{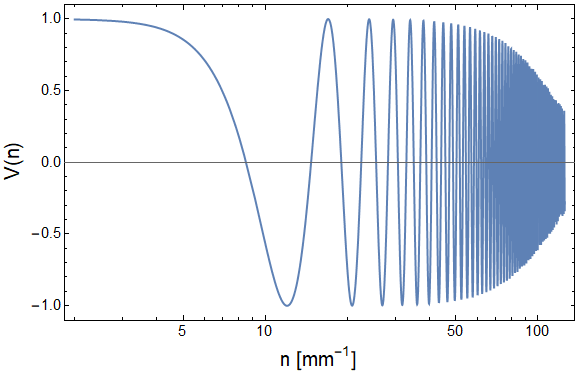}
(c)\includegraphics[width=0.45\textwidth]{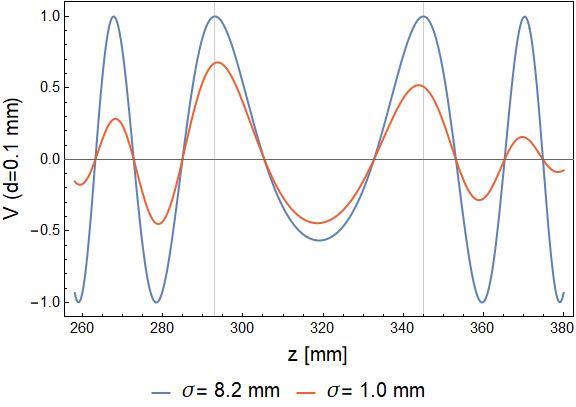}
(d)\includegraphics[width=0.45\textwidth]{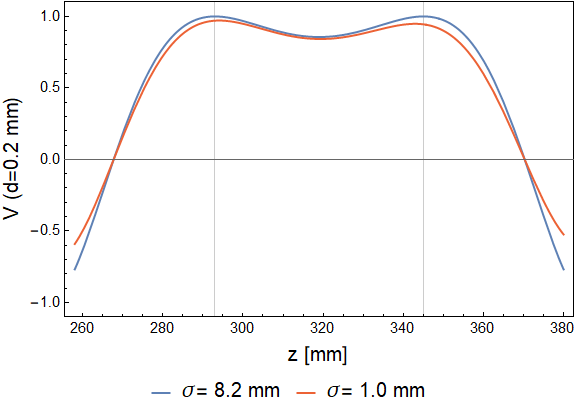}
\caption{Panel (a) shows the visibility \eqref{eq:vis} of the CPI-AP image of a pattern, as a function of its period $d$ obtained for $\lambda=532\,\mathrm{nm}$, $z_a=293\,\mathrm{mm}$, $z_b=345\,\mathrm{mm}$, $\sigma=8.2\,\mathrm{mm}$, with the object placed in the axial position $z=z_m=(z_a+z_b)/2$. Panel (b) shows the same quantity plotted in terms of the spatial frequency of the sample, expressed as the number of peaks per unit length $n=1/d$. Panels (c) and (d) show the plots of the visibility $V(d)$ as a function of the axial position $z$ of the pattern for two fixed periodicity values $d=0.1\,\mathrm{mm}$ and $d=0.2\,\mathrm{mm}$. In both cases, the blue and orange curves correspond to the visibility obtained for lens width $\sigma=8.2\,\mathrm{mm}$ and $\sigma=1\,\mathrm{mm}$, respectively, while other parameters are fixed as in panels (a)-(b). }\label{fig:visibility}
\end{figure} 

In order to characterize the resolution for CPI-AP with respect to standard imaging, we first define the visibility of the main peaks of a periodic object with period $d$, as
\begin{equation}\label{eq:vis}
V(d)=\frac{\Sigma_{\mathrm{ref}}(0;d)-\Sigma_{\mathrm{ref}}(d/2;d)}{\Sigma_{\mathrm{ref}}(0;d)+\Sigma_{\mathrm{ref}}(d/2;d)} = \frac{C_1(d)}{C_0(d)+C_2(d)},
\end{equation}
with $0$ and $d/2$ corresponding to the values of the $x_r$ coordinate at which the object intensity profile reaches a pair of adjacent maximum and minimum, respectively. Notice that we omit the irrelevant dependence on $y_r$ in the above equation.
In Figure \ref{fig:visibility} we show the plot of $V(d)$ calculated through Eq.~\eqref{eq:vis} for the aforementioned experimental parameters. We notice that for small values of $d$ the visibility curve shows a strongly oscillating behavior. This is due to the peculiar nature of resolution loss in this imaging system, in which nonlinearity entails the appearance of secondary peaks in between the object spatial periodicity, as shown in Figure~\ref{fig:object}. 
In particular, the negative parts of Figure~\ref{fig:visibility}(a) represent cases where the secondary peaks become dominant over the primary ones,  corresponding to the maxima of the actual object profile. The infinite revivals of high (and positive) visibility for small values of $d$ are due to the spatial indefiniteness of the pattern, and are not relevant in practice for imaging of finite-size objects. Revivals out of the expected high-visibility regions around $z_a$ and $z_b$ can be observed also in Figure~\ref{fig:visibility}(c), representing the visibility as a function of the pattern axial position for $d=0.1\,\mathrm{mm}$. Notice that a smaller lens aperture tends to damp oscillations far from the focused planes. The plot in Figure~\ref{fig:visibility}(d) shows a qualitatively different situation, in which a range without discontinuities, enclosing both $z_a$ and $z_b$, is characterized by visibility above the considered limit.

\begin{figure}
\centering
\includegraphics[width=0.7\textwidth]{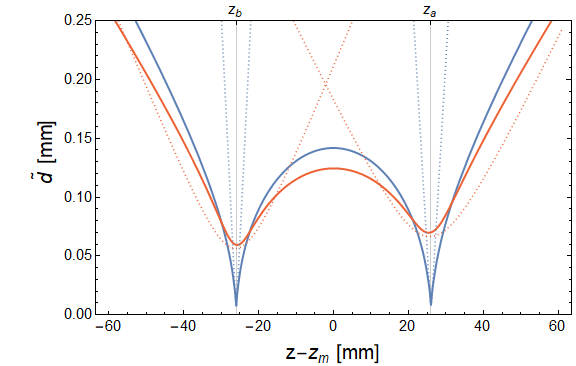}
\includegraphics[width=0.7\textwidth]{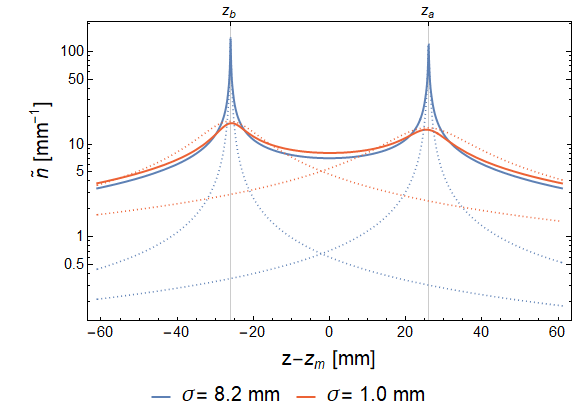}
\caption{\textit{Upper panel}: Resolution limits $\tilde{d}$, determined according to the definition in Eq.~\eqref{eq:resolution_condition} with $V_{\lim}=0.15$. The two solid lines represent the cases in which the standard deviation of the effective lens aperture is equal to $8.2\,\mathrm{mm}$ (blue line) and $1\,\mathrm{mm}$ (red line). The dotted lines in the same colors report, for comparison, the resolution limit of the standard images collected by the two detectors. \textit{Lower panel}: Resolution limits in terms of spatial frequency $\tilde{n}=1/\tilde{d}$. Style and colors of lines have the same meaning as in top panel. 
Notice that in the considered setup the resolutions in the two planes focused on the detectors are different, since the lens aperture is seen at different angles from each of the two planes. All plots are obtained for $\lambda = 532\,\mathrm{nm}$.}\label{fig:resolution}
\end{figure}

Though it is tempting to assume that a pattern is ``resolved'' whenever the visibility (Eq.~\eqref{eq:vis}) of the main peaks is larger than a certain limiting value, the definition of a proper resolution limit must also take into account the peculiar emergence of the second harmonic. In the case reported in Fig.~\ref{fig:object}(c), for example, the secondary peaks are artefacts characterized by such a good contrast that they could be easily interpreted as object features. To address this problem, we define the resolution limit by means of the available analytical results of the image. In particular, considering that the minimum value of the image in between the two peaks reads 
\begin{equation}
    \Sigma_{\mathrm{ref},\min}(d) = F(d) \left[ C_0(d) - C_2 (d) - \frac{C_1^2(d)}{8 C_2(d)} \right],
\end{equation}
we can express the visibility of the secondary peaks as:
\begin{equation}
    V_2(d) = \frac{\Sigma_{\mathrm{ref}}(d/2;d) - \Sigma_{\mathrm{ref},\min}(d) }{\Sigma_{\mathrm{ref}}(d/2;d) + \Sigma_{\mathrm{ref},\min}(d) } = \frac{[C_1(d) - 4 C_2(d)]^2}{8C_2(d)[2 C_0(d) - C_1(d)] - C_1^2(d) }.
\end{equation}
The resolution limit $\tilde{d}$ can now be defined as the value of the input pattern period $d$ such that, given a visibility limit $V_{\lim}$, the object pattern is resolved for $d>\tilde{d}$, while the secondary peaks are not, namely:
\begin{equation}\label{eq:resolution_condition}
    V(d) > V_{\lim} > V_2(d) \qquad \text{for all } d>\tilde{d}.
\end{equation}
The value of $V_{\lim}$ is a choice of the user, who can be interested in defining a tighter or looser visibility criterion, according to the application. It is worth noticing that, in any case, the class of resolution criteria $V>V_{\lim}>V_2$ automatically rules out the possibility for an image like the one in Figure~\ref{fig:object}(c), characterized by $V_2>V$, to be considered as nominally resolved, regardless of the chosen value of $V_{\lim}$. On the other hand, the image in Figure~\ref{fig:object}(b), characterized by $V=0.47$ and $V_2=0.15$, could be considered resolved in the cases $0.15<V_{\lim}<0.47$, while the one in Figure~\ref{fig:object}(a), with $V\simeq 1$ and $V_2\simeq 0$, is resolved for a much wider range of $V_{\lim}$ values.

In Figure~\ref{fig:resolution}, we plot the resolution limit obtained by the condition \eqref{eq:resolution_condition} with $V_{\lim}=0.15$. It is worth noticing that, despite in previous literature the visibility of the secondary peaks in between the slit images was neglected, the reported quantity shows consistent behavior with that obtained in Refs. \cite{cpiap,resol_limit}, with minima close to the two reference planes $z=z_j$, and a local maximum close to $z=(z_a+z_b)/2$. We further observe that, by changing the width of the pupil function (by modifying $\sigma$), the curve of the resolution limit changes: the blue line is obtained with $\sigma=8.2\,\mathrm{mm}$, while the red line corresponds to the case of a smaller pupil function ($\sigma=1.0\,\mathrm{mm}$).
Motivated by these observations, we calculate an analytic expression for visibility of a periodic object in the limit of a pupil function with an infinite width. For $\sigma\to\infty$, we obtain the following expressions:
\begin{align}
    V_{\infty}(d) & = \lim_{\sigma\to\infty} V(d) =\cos\left[\frac {2\pi^2  (z - z_a) (z - z_b)} {d^2 k (z_a - z_b)} \right], \\
    V_{2,\infty} (d) & = \frac{1 - V_{\infty}(d) }{3 + V_{\infty} (d) } . 
\end{align}
Evidently, the visibility $V_{\infty}$ of the main peaks approaches unity (also implying $V_{2,\infty}=0$) in the focused cases ($z=z_{j}$) for all $d$ values, and in the case $d\to\infty$ or $k\to\infty$, for all $z$ values. Instead, it vanishes for
\begin{equation}
    d=d_m= \sqrt{\frac{2 \lambda}{(2m+1)}
    \left\lvert \frac{\left(z-z_{a}\right)\left(z-z_{b}\right)}{\left(z_{a}-z_{b}\right)}\right\rvert}
\end{equation}
with $\lambda=2\pi/k$, $m=0,1,2...$ and $d_0$ being the largest value. If the $\sigma\to\infty$ limit is valid, the periodicity $\tilde{d}$ corresponding to the visibility limit in Eq.~\eqref{eq:resolution_condition} can be analytically determined as
\begin{equation}\label{eq:limhighsigma1}
    \tilde{d} \simeq \sqrt{ \frac{\pi \lambda}{\phi(V_{\lim})} \left\lvert \frac{\left(z-z_{a}\right)\left(z-z_{b}\right)}{\left(z_{a}-z_{b}\right)}\right\rvert  } ,
\end{equation}
with
\begin{equation}\label{eq:limhighsigma2}
    \phi(V_{\lim}) = \left\{ \begin{array}{ll} \displaystyle
         \arccos\left( \frac{1-3V_{\lim}}{1+V_{\lim}} \right) & \text{for } 0\leq V_{\lim} \leq \sqrt{5}-2 \\  \displaystyle
         \arccos(V_{\lim}) & \text{for } \sqrt{5}-2 < V_{\lim} \leq 1
    \end{array} \right.
\end{equation}
(where $\sqrt{5}-2=0.236$). Such an estimate of $\tilde{d}$ provides an excellent approximation out of the natural DOF of the two focusing planes. Hence, as expected, it has a wider range of validity for the higher value of $\sigma$ (blue curve). Referring to Figure~\ref{fig:resolution}, it is worth noticing that, while for $\sigma=8.2\,\mathrm{mm}$ (blue curves) the criterion $V_2(d) < V_{\lim}=0.15$ is dominant, as expected from Eqs.~\eqref{eq:limhighsigma1}-\eqref{eq:limhighsigma2}, the resolution limit for $\sigma=1.0\,\mathrm{mm}$ (red curves) is determined by $V(d) > V_{\lim}=0.15$. This occurs because lower apertures tend to increase blurring in the refocused images, thus reducing the visibility of the secondary peaks.

For comparison, we also report in Figure~\ref{fig:resolution} the resolution limits at $V_{\lim}=0.15$ of the first-order images 
\begin{equation}
    I_j(\br_j) = \int \de^2\br_o A(\br_o) \left\lvert g_j(\br_j,\br_o) \right\rvert^2 \qquad (j=a,b)
\end{equation}
captured by each of the two detectors separately. In this case, since the visibility of the image pattern reads
\begin{equation}
    V_j (d) = \exp \left\{ - \left( \frac{\pi z}{d k \sigma} \right)^2 \left[ 1 + \left( \frac{k\sigma^2}{z}\right)^2 \left( 1 - \frac{z}{z_j} \right)^2 \right] \right\} \qquad (j=a,b),
\end{equation}
the resolution limit has the following dependence on the system parameters:
\begin{equation}
    \tilde{d}_j = \frac{\lambda z}{2\sigma \sqrt{\lvert\ln(V_{\lim})\rvert}} \sqrt{1 + \left(\frac{2\pi\sigma^2}{\lambda z}\right)^2 \left( 1 - \frac{z}{z_j} \right)^2 }  \qquad (j=a,b).
\end{equation}
From the results in Figure~\ref{fig:resolution}, we observe that the resolution of CPI is improved with respect to that of the standard images in practically all the considered range in the case $\sigma=8.2\,\mathrm{mm}$, roughly corresponding to a lens of diameter around $30\,\mathrm{mm}$. Instead, a lens with $\sigma=1\,\mathrm{mm}$ provides a standard-image resolution that is everywhere in the represented range comparable, if not better, than the asymptotic resolution of CPI. However, such an apparent advantage comes at the expense of a heavy loss of resolution in the focused planes, by a factor larger than 8, and does not entail the possibility of viewpoint choice and three-dimensional imaging, which is inherent to plenoptic imaging techniques.

\section{Discussion}\label{sec:discussion}

We have defined and characterized the resolution limits in CPI-AP for objects with periodic intensity profile in space. The difficulty in defining resolution limits in an unambiguous way has clearly emerged, since the visibility of the periodic image obtained from the calculation shows an oscillating behavior when the object spatial frequency increases.

The definition of resolution limits in the present work is conceptually analogous to the one considered in previous literature on the topic, which was based on the ability to discriminate a double slit with very specific features, namely, a center-to-center distance equal to twice the slit width (see, e.g., Ref.~\cite{cpiap} for the CPI-AP setup and Ref.~\cite{cpi_exp} for a different CPI system). In that case, the center-to-center distance plays the role of a carrier-wave periodicity of the object. However, in the present analysis, we have strengthen the definition of the resolution limit by requiring that the secondary peaks, which tend to appear in between each pair of object features, cannot be misinterpreted as object parts, according to the proposed visibility criterion. Despite this difference, the obtained results are qualitatively consistent with previous literature in terms of the variation of the resolution with varying object axial position. We remark that, though CPI-AP performs better than both standard imaging and classical plenoptic imaging in terms of resolution, a full evaluation of the advantages with respect to standard techniques must also take into account the problem of noise, which affects correlation imaging in a specific way (see, e.g., Refs.~\cite{cpi_snr,CPI_SNR_1}). A thorough discussion of this issue will be matter for future research.

Despite the nonlinearity of the CPI image, our results demonstrate that an object with a periodic pattern offers more insight in the performance of a CPI-AP setup, also from a qualitative point of view. Furthermore, by observing the behavior of the resolution limits obtained with different realistic widths of the pupil function, we can state that, in experimentally accessible conditions, the resolution limits are essentially independent of the size of the lens, and well approximated by the infinite-lens-size expression. We finally remark that, though the presented results are obtained in the specific setup of CPI-AP, they can be easily generalized to different CPI configurations. In fact, the main difference of CPI-AP with respect to other CPI protocols, is the fact that, in the latter, one of the two reference planes has a well-defined position (e.g., coinciding with the principal plane of a lens) instead of being arbitrary.

\backmatter

\bmhead{Authors' contributions}
Conceptualization, B.S., F.V.P., M.D.; methodology, F.S., G.M., F.V.P.; software, F.S.; validation, all authors; formal analysis, F.S.; investigation, F.S., G.M.; writing---original draft preparation, F.S.; writing---review and editing, all authors; visualization, F.S.; supervision, F.V.P. and M.D.; project administration, M.D.; funding acquisition, M.D., B.S. F.S. All authors have read and agreed to the published version of the manuscript.

\bmhead{Funding}
F.S. is supported by Research for Innovation REFIN - Regione Puglia POR PUGLIA FESR-FSE 2014/2020; G.M., F.V.P and M.D. are supported by Istituto Nazionale di Fisica Nucleare (INFN) through project PICS4ME. G.M, B.S., F.V.P and M.D. are supported by project Qu3D, funded by the Italian Istituto Nazionale di Fisica Nucleare, the Swiss National Science Foundation (grant 20QT21 187716 “Quantum 3D Imaging at high speed and high resolution”), the Greek General Secretariat for Research and Technology, the Czech Ministry of Education, Youth and Sports, under the QuantERA programme, which has received funding from the European Union's Horizon 2020 research and innovation programme. M.D. is supported by European Union-NextGenerationEU PE0000023 - ``National Quantum Science and Technology Institute''. F.V.P. is supported by European Union-NextGenerationEU CN00000013 - ``National Centre for HPC, Big Data and Quantum Computing''.

\bmhead{Data Availability Statement}
This manuscript has no associated data.


\end{document}